
\font\asi = msbm7 scaled\magstep0
\font\asd = msbm10 scaled\magstep0
\font\eightrm = cmr8
\font\sixrm = cmr6
\font\large=cmr12 scaled \magstep2
\font\sixi = cmmi6
\font\eighti = cmmi8
\input harvmac
\def\an#1{\hbox{\asd #1}}
\def\ai#1{\hbox{\asi #1}}

\lref\n{J. Neukirch, {\it Algebraische Zahlentheorie}, Springer-Verlag, Berlin
1992.}

\lref\5{E.W. Barnes, Q. J. Pure Appl. Math, {\bf 31}, 264, (1900); J. Milnor,
L'Enseignement Math. {\bf 29}, 281, (1983).}

\lref\3{M.J. Ablowitz and H. Segur, {\it Solitons and the Inverse Scattering
Transform}, SIAM, Philadelphia, 1981.}

\lref\a{ M.A. Olshanetsky and A.M. Perelomov, Phys. Rep. {\bf 94}, 313 (1984);
R.F. Wehrhahn, Rev. Math. Phys. {\bf 6}, 1339 (1994).}

\lref\b{ P.G.O. Freund, Phys. Lett. B {\bf 257}, 119 (1991);
L. Brekke and P.G.O. Freund, Phys. Rep. {\bf 233}, 1 (1993).}

\lref\c {L.O. Chekhov, J. Math. Phys. {\bf 36}, 414 (1995).}

\lref\d {L.D. Faddeev and B.S. Pavlov, Sem. LOMI 27, 161 (1972);
P.D. Lax and R.S. Phillips, {\it Scattering Theory for Automorphic Functions},
Princeton University Press, Princeton, N.J. 1976. }

\lref\e {S. Gelbart and F. Shahidi, {\it Analytic Properties of Automorphic
L-Functions}, Academic Press, Inc. Boston, 1988.}

\rm
\par\noindent
{\rightline {\vbox{\hbox{hep-th/9507031}
                   \hbox{EFI 95-40}
                   \hbox{June 1995}}}}
\bigskip\noindent
\bigskip\noindent
\centerline{{\large Adelic Integrable Systems}$^{\raise3pt\hbox{$\scriptstyle
1$}}$}
\vskip 1.0cm
\centerline{Mircea Pigli}
\vskip 1.0cm
\centerline{ \it Enrico Fermi Institute and Department of Physics,}
\centerline{ \it University of Chicago, Chicago, IL 60637, USA}
\vskip 1.0cm
\centerline{\bf Abstract}
\medskip\noindent
Incorporating the zonal spherical function (zsf) problems on real and $p$-adic
hyperbolic planes into a Zakharov-Shabat integrable system setting, we find a
wide class of integrable evolutions which respect the number-theoretic
properties of the zsf problem. This means that at {\it all} times these real
and $p$-adic systems can be unified into an adelic system with an $S$-matrix
which involves (Dirichlet, Langlands, Shimura...) L-functions.
\vfill
\hrule
\vskip 3pt
$^1$ Work supported in part by NSF Grant PHY-91-23780
\footline={\hfil}
\eject

\newsec{Introduction}
\footline={\hss\tenrm\folio\hss}
Scattering theory on real {\a} and $p$-adic {\b} symmetric spaces can be
unified
in an adelic context {\b}, {\c}. This has the virtue of producing $S$-matrices
involving the Riemann zeta function and of throwing new light on earlier
work {\d} concerning
scattering on the noncompact finite-area fundamental domain of $SL(2,${\an Z})
on the real hyperbolic plane $H_\infty$.

The real hyperbolic plane is a smooth manifold and as such quantum mechanics
on $H_\infty$ involves a second order Schr\"{o}dinger {\it differential}
equation. By contrast the $p$-adic hyperbolic planes $H_p$ are discrete
spaces (trees), and the corresponding Schr\"{o}dinger equations are second
order {\it difference} equations. The Jost functions, and therefore the
$S$-matrices from all these {\it local}
problems combine in adelic products, which then involve the Riemann zeta
function {\b}.

At a given initial time consider all these (``$S$-wave'') scattering problems
and then let
all of them undergo an {\it integrable} time evolution. In general such an
evolution need not respect the number-theoretic endowment of the initial
problem. In other words, even though at the initial time the real and $p$-adic
scattering problems assembled into an interesting adelic scattering problem,
at later times
this need no longer be so. We want to explore here the conditions under which
the integrable evolution respects adelizability and to see what kind of
scattering problems can be obtained this way at later times. Specifically, we
will incorporate the initial scattering problem into a Zakharov-Shabat (ZS)
system and follow its integrable evolution. For the $p$-adic problems, time
has to be discrete and for adelic purposes time then has to be discrete in the
real problem as well. We will see that adelic products can be meaningful
at later times in the evolution of such a system and that along with the
Riemann zeta function involved in the adelic problem at the initial time,
various (Dirichlet, Langlands, Shimura,...) $L$-functions {\e} appear at later
times.

\newsec{Integrable Evolution of the $p$-adic Zonal Spherical Function
Problem}

As mentioned in the introduction, we consider ``$S$-wave'' scattering
problems on local (real and $p$-adic) hyperbolic planes and embed them in
integrable ZS systems. We start by setting up quantum mechanics on these
hyperbolic planes and then finding $S$-wave solutions (i.e. solutions
independent on the angular
variable) thereof. Dealing with $S$-waves corresponds mathematically to
the zonal spherical function (zsf) problems on these hyperbolic
planes. We first consider the $p$-adic hyperbolic plane
$H_p=SL(2,$\an Q$_p)/SL(2,$\an Z$_p)$. This $H_p$ is a discrete space, a
homogeneous Bruhat-Tits-Bethe tree of incidence number $p+1$ (or equivalently,
branching number $p$) and the radial coordinate is an integer, say $n$. The zsf
$w_n$ on $H_p$ solves the familiar difference equation {\b}
$$p w_{n+2} - \sqrt{p} \bigl(p^{{ik}\over 2}+p^{-{{ik}\over 2}}\bigr) w_{n+1} +
w_n = 0.    \eqno(2.1)$$
It is this equation that we wish to embed in an integrable ZS system. We do
this as follows. Consider the two-component ZS system governed by the
equations
$$\eqalign{u_{n+1} &= z u_n +Q_n v_n \cr
                  v_{n+1} &= P_n u_n + z^{-1} v_n \cr}.     \eqno(2.2)$$
By decoupling this set of first order difference equations, one obtains
separate second order difference equations for the two components of the
ZS doublet. In particular $v_n$ obeys
$${{P_n}\over{P_{n+1}}(1-P_nQ_n)}v_{n+2}
-{{1}\over{1-P_nQ_n}}\biggl[{{P_n}\over{P_{n+1}}}z^{-1} + z\biggr]v_{n+1} +v_n
=0.      \eqno(2.3)$$
We achieve the embedding of the zsf equation (2.1) in the ZS system, by
requiring that this $v_n$ essentially reproduce the zsf $w_n$, specifically
that for some real $\nu$
$$w_n = p^{-\nu n} v_n. \eqno(2.4)$$
{}For this to be the case, $Q_n$ and $P_n$  must obey the relations

$$\eqalign{&{{P_n}\over{P_{n+1}}(1-P_nQ_n)}p^{2\nu} = p    \cr
&{{1}\over{1-P_nQ_n}}\biggl[{{P_n}\over{P_{n+1}}}z^{-1} + z\biggr]p^{\nu} =
\sqrt{p} \bigl(p^{{ik}\over 2}+p^{-{{ik}\over 2}}\bigr).\cr }      \eqno(2.5)
$$
Here we must require $Q_n$ and $P_n$ to be $z$-independent and the
relation between $z$ and $k$ to be $n-$independent. These requirements
result in
$$z = p^{{ik-\rho}\over 2} \eqno(2.6a) $$
where for convenience we introduced the new parameter
$$\rho = 2\,\nu -1  \eqno(2.6b)$$
with $\nu$ as in equation (2.4).

On account of the invariance of equation (2.1) under a sign change for $k$,
 in equation (2.6) $k$ can just as well be replaced by $-k$. We opt for the
relation (2.6a), as it stands. When used in eq.(2.5) it gives

$${{P_{n+1}}\over{P_n}}=p^\rho, \quad Q_n = 0  \eqno(2.7)$$
so that

$$P_n = \sigma_p (0)\,p^{\rho n} \quad {\rm and} \quad Q_n = 0, \eqno(2.8)  $$
where $\sigma_p (0)$ is an integration constant.

This determines, via (2.1) and (2.4)

$$\eqalign{u^0_n &= p^{{(ik-\rho)n}\over 2}\,u^0_0   \cr
 v^0_n &= {{p}\over{p+1}} \biggl[p^{{(ik+\rho)n}\over 2}\,c^0(k;p) +
p^{{(-ik+\rho)n}\over 2}\,c^0(-k;p) \biggr] \cr}  \eqno(2.9) $$
\noindent
where $c^0(k;p)$ is the Jost-Harish-Chandra function for the scattering
problem on a tree \b,
$$c^0(k;p) ={{\zeta_p(ik)}\over{\zeta_p(ik+1)}}
={{1-p^{-ik-1}}\over{1-p^{-ik}}} \eqno(2.10)$$
\noindent and
$$u^0_0 ={{ p^{{ik+\rho+2}\over 2} - p^{{-ik+\rho}\over 2}}
\over{\sigma_p(0)(p+1)}} \eqno(2.11)$$

We now wish to find an integrable  time-evolution for this system. We will
 assume, throughout this paper, that time is discrete and its value will be
indicated by a superscript, say $m$. Imposing the condition that at any later
time the spatial dependence of the system should still be of the ZS type, leads
to a compatibility relation between the time and space evolutions.
Let us introduce matrix notation

$$\Psi^m_n = \pmatrix{u_n^m\cr \noalign{\smallskip} v^m_n\cr}\quad R^m_n =
\pmatrix{z& Q^m_n\cr \noalign{\smallskip} P^m_n& z^{-1}\cr } \eqno(2.12a)$$
where as explained the superscript $m$ indicates time and the subscript $n$
distance to
the origin of the tree. The ZS equations at time $m$ are then
$$ \Psi^m_{n+1} = R^m_n \Psi^m_n \eqno(2.12b).$$

Introducing the time evolution matrix
$$ M^m_n = \pmatrix {A^m_n& B^m_n\cr \noalign{\smallskip} C^m_n& D^m_n\cr},
\eqno(2.12c)$$
the time evolution of the system is governed by the equation
$$\Psi^{m+1}_n = M^m_n \Psi^m_n. \eqno(2.12d)$$

The compatibility condition between the time evolution and the eigenvalue
problem is
$$R^{m+1}_n M^m_n = M^m_{n+1} R^m_n. \eqno(2.13) $$

In order to solve equations (2.11) we need to make some further assumptions, as
otherwise we
 have six unknowns and four relations. We will expand the elements of the
time-evolution matrix in powers of $z$, and retain only the first term.
However, because of the way $z$ and $z^{-1}$ occur in $R^m_n$, these expansions
will  be
chosen differently \3 for the four matrix elements of $M^m_n$:

$$\eqalign{A^m_n & =  a^m_n + z^{-2} \alpha^m_n\cr
           B^m_n & = zb^m_n + z^{-1} \beta^m_n\cr
}\qquad\eqalign{
           C^m_n & = zc^m_n + z^{-1} \gamma^m_n\cr
           D^m_n & = d^m_n + z^2 \delta^m_n\cr}. \eqno(2.14)  $$

As shown in detail in Appendix A, these assumptions lead to the following
solution for the ZS system (2.12):
$$R^m_n = \pmatrix{ z&0\cr
\noalign{\smallskip}\sigma_p(m)p^{\rho (n-m)} &z^{-1}\cr} \eqno(2.15a)$$
\noindent and
$$M^m_n = \pmatrix{a(m)+ z^{-2}\,\alpha(m) &0 \cr
\noalign{\smallskip} -p^{\rho(n-m-1)}\bigl[\sigma_p(m)\delta(m)z +
\sigma_p(m+1)\alpha(m)z^{-1}\bigr] & d(m) + z^2\delta(m)\cr} \eqno(2.15b)$$

\noindent where $\sigma_p(m)$ is an arbitrary function of the discrete time $m$
and $ a(m)$, $\alpha(m)$, $d(m)$ and $\delta(m)$ are equal respectively to
$a_0^m$, $\alpha_0^m$, $d_0^m$ and $\delta_0^m$ of eq.(2.14) and must be
related by:
$$\sigma_p(m+1)\bigl[\alpha(m)\,p^\rho +a(m)\bigr] = \sigma_p(m)\bigl[\delta(m)
+ d(m)\,p^\rho\bigr] \eqno(2.15c)$$

Just like $\sigma_p(m)$, the quantities $a(m)$, $\alpha(m)$, $d(m)$ and
$\delta(m)$ all depend, of
course, on the Bethe lattice branching number $p$, but we choose not
to explicitly indicate this dependence.

Corresponding to this solution, the ZS doublet, which at time $m=0$ takes the
form (2.9), evolves into:
$$ \eqalign{u^m_n &=  u^0_n \prod_{j=0}^{m-1}{[a(j) + p^{\rho
-ik}\alpha(j)]}\cr            v^m_n &=  {p\over{p+1}}p^{{\rho n}\over
2}\Bigl[p^{{ikn}\over2}c^0(k;p)
{{\sigma_p(m)}\over{\sigma_p(0)}}\prod^{m-1}_{j=0}{[a(j)p^{-\rho}+\alpha(j)p^{-ik}]} + \cr &\hskip 0.7in +p^{-{{ikn}\over 2}}c^0(-k;p)\prod^{m-1}_{j=0}{[d(j) + p^{ik-\rho}\delta(j)]}\Bigr].\cr} \eqno(2.16)$$
with $u^0_n$ given by eq. (2.9). For the S-matrix at $m\ne 0$ to be unitary,
in the expression for $v^m_n$,
$p^{{ikn}\over2}$ and $p^{-{{ikn}\over2}}$ must have complex conjugate
coefficients, which requires that equation (2.15c) be replaced by the stronger
pair  of equations:
$$ \eqalign{\sigma_p(m+1)\, \alpha(m)\,p^\rho &= \sigma_p(m)\,\delta(m)\cr
           \sigma_p(m)\,d(m)\,p^\rho &= \sigma_p(m+1)\,a(m),\cr} \eqno(2.15d)$$
which, of course, imply eq.(2.15c).

With these relations one can eliminate $a(m)$ and $\alpha(m)$ and the
formulae for $u^m_n$ and $v^m_n$ simplify considerably. To see this,
it is convenient to renormalize $u$ and $v$ by removing from them an
overall factor which depends {\it only} on the discrete
time $m$. This way, we introduce the new functions $\tilde u^m_n$ and
$\tilde v^m_n$ as follows:
$$\eqalign{\tilde u^m_n &= f(m)\,u_n^m\cr
           \tilde v^m_n &= g(m)\,v_n^m\cr} \eqno(2.17)$$
\noindent with
$$\eqalign{f(m) &= \prod_{j=0}^{m-1} {d^{-1}(j)}\cr
           g(m) &= {\sigma_p(m)\over\sigma_p(0)}\,f(m)\cr} \eqno(2.18)$$

Then, the final simplified  form of $\tilde u^m_n$ is
$$\tilde u_n^m = \Lambda (m,\chi(p),k;p)u^0_n \eqno(2.19)$$

\noindent and $\tilde v_n^m$ is given by the second equation (2.9) with
the Jost-Harish-Chandra $c-$function $c^0(k;p)$ replaced by
$$c^m(k;p) = c^0(k;p)\,\Lambda(m,\chi(p),k\,;p). \eqno(2.20)$$
In equations (2.19) and (2.20), the function $\Lambda$ is given by
$$\Lambda(m,\chi(p),k\,;p) = \prod_{j=0}^{m-1}{\bigl[1 -
\chi^{j}(p)^{-ik}\bigr]} \eqno(2.21)$$
\noindent with
$$\chi^{j}(p) = -{\delta(j)\over d(j)}\,p^{-\rho}. \eqno(2.22)$$
In other words $\tilde u_n^m$ and the $c-$function both accrete the
same factor $\Lambda$.

In the tendentious notation just used, it is clear that
$\Lambda$ is a candidate factor in an Euler product, provided only
$\chi^{j}(p)$ has a ``nice'' $p-$dependence. By this we mean that in the
simplest case $\chi^{j}(p)$ is a Dirichlet character, or something
similar. We shall return to this point when we treat the adelic
problem.

Finally, $\tilde u^m_n$ does not correspond to a scattering problem, since
it does not contain an incoming wave. Yet, even $\tilde u^m_n$ has encoded
in it an object $l^m(k,p)$, which in adelic considerations will play a
role similar to that played by the Jost-Harish-Chandra function $c^m(k,p)$,
which is encoded in the $v$'s. This $l^m(k,p)$ is found by rewriting
$\tilde u^m_n$ in the form
$$
\tilde u^m_n=l^m(k,p){p^{ik(n+1)-\rho(n-1)+2\over 2}\over (p+1)\sigma_p(0)}.
\eqno(2.23)
$$
Comparing with equations (2.19), (2.9)-(2.11), we then find
$$
l^m(k,p)={\Lambda(m,\chi(p),k;p)\over 1-p^{-(ik+1)}}.
\eqno(2.24)
$$

Next we have to deal with the place at infinity, which involves the
real hyperbolic plane $H_\infty=SL(2,{\an Q_\infty})/SO(2,{\an Q_\infty})$
where ${\an Q_\infty} \equiv {\an R}$ is the field of real numbers.

\newsec {Integrable evolution of the real Zonal Spherical Function Problem}
The adelic partner of the zsf problems on the $p$-adic hyperbolic planes
tackled in section 2, is the zsf problem on the real hyperbolic plane $H_\infty
= SL(2,{\an R})/SO(2,{\an R})$. The corresponding eigenvalue equation for the
radial Laplacian is:

$$w''+ 2\coth 2x \,w' + (k^2+1) \,w = 0 \eqno(3.1) $$

This is no longer a difference equation, but rather a differential
equation, $( w' \equiv {\rm d}w/{\rm d}x)$,
since, unlike its $p$-adic counterparts, the real hyperbolic plane is a
continuous manifold and not a discrete Bruhat-Tits-Bethe tree. We therefore
encounter a continuous radial coordinate $x$, while time $m$ in the ZS system
must stay discrete, for a proper match with the $p$-adic cases.
The standard ZS problem in this case is {\3}
$$\eqalign{(u^0)' &= i\zeta u^0 + q^0 v^0 \cr
                   (v^0)' &= P^0 u^0 - i\zeta v^0 \cr}  \eqno(3.2) $$
with  $\zeta$  the counterpart of the spectral variable $z$. Specifically,
$\zeta$ and $z$ are related as
$$ z = {\rm e}^{i\zeta}. \eqno(3.3) $$

Were we to deal with this problem, we could easily find its solution and
integrable time evolution. Yet expanding in $\zeta$ {\it not} being the same as
expanding in $z$ in section 2, the nice adelic match would be lost. Therefore,
we shall consider a different, adelically better suited alternative, namely

$$\eqalign{(u^0)' &= z\,u^0 + Q^0(x)\,v^0\cr
           (v^0)' &= P^0(x)\,u^0 + z^{-1}\,v^0 \cr}. \eqno(3.2') $$

We will return to the original ZS problem (3.2) in Appendix C.

Just like in the discrete case, the second order
equation obeyed by $v^0$ is

$$(v^0)'' - (v^0)'\Bigl[z+z^{-1}+{{(P^0)'}\over{P^0}}\Bigr] + v^0
\Bigl[1-Q^0P^0 +z^{-1}{{(P^0)'}\over{P^0}}\Bigr] = 0 \eqno(3.4) $$
which, after a 'gauge' transformation similar to (2.3),

$$v^0(x) = w^0(x)\tau(x) \eqno(3.5)$$
upon comparison with (3.1) yields the conditions

$$\eqalign{2\coth 2x &= 2{{\tau'}\over{\tau}} -z - z^{-1} -
{{(P^0)'}\over{P^0}} \cr
k^2+1 &= {{\tau''}\over{\tau}} - {{\tau'}\over{\tau}}\Bigl[z +
z^{-1}+{{(P^0)'}\over{P^0}}\Bigr] + 1 - P^0 Q^0
+z^{-1}{{(P^0)'}\over{P^0}}  \cr} \eqno(3.6) $$

\noindent where $P^0$ and $Q^0$ are $z$-independent, and the relation between
$z$ and $k$ has to be $x$-independent. Once these conditions are
imposed, it follows that
$$\eqalign{P^0 =& \  \sigma(0) {\rm e}^{2\nu x}\cr
Q^0 =& -{1\over{\sigma(0)\sinh^2 2x}}{\rm e}^{-2\nu x} \cr} \eqno(3.7)$$
and
$$\eqalign{ \tau(x) =& \tau_0(x)(\sinh 2x)^{1/2}{\rm e}^{(\nu
+{{z+z^{-1}}\over2})x} \cr z - z^{-1} =& -2\nu \pm 2ik \cr }\eqno(3.8)$$
with $\nu$ an arbitrary complex number. Again, everything is symmetric with
respect to changing the sign of $k$, and we chose the + sign
in (3.8). The $x$-dependence in relations (3.7) shows that at large $x$
only one of the two functions, $P^0$ and $Q^0$,
survives, depending upon the real part of $\nu$. In what follows,
without any loss of generality, we assume ${\rm Re }\,\nu \ge 0$, so as to
preserve the resemblance with
the discrete case.

The $x \rightarrow \infty$ asymptotic form of the solution to (3.1) is

$$\eqalign{ u^0(x) &\sim
2ik\,{{\tau(x)}\over{\sigma(0)}}c^0\,(k;\infty){\rm
e}^{(ik-1-2\nu)x}\cr
v^0(x) &\sim \tau(x)\bigl[c^0\,(k;\infty)\,{\rm e}^{(ik-1)x}
+ c^0\,(-k;\infty)\,{\rm e}^{-(ik+1)x}\bigr]\cr }\eqno(3.9)$$
where
$$c^0\,(k;\infty) =
\pi^{-1/2}{{\Gamma\bigl[{1\over2}ik\bigr]}\over{\Gamma\bigl[{1\over2}(ik+1)\bigr]}} = {\zeta_\infty(ik)\over\zeta_\infty(ik+1)}\eqno(3.10)$$
Next, we need to find an integrable time evolution for the system that
is compatible with (3.1). The compatibility condition in this case is

$$(M^m)' = R^{m+1}M^m - M^mR^m \eqno(3.11)$$
and the superscript $m$ again stands for discrete time.
We shall solve equation (3.11) subject to the condition that at {\it
all} times $m$, the asymptotic $R$ matrix at large radial distances
$x$ have the form

$$R^m \sim \pmatrix{z & 0\cr \noalign{\smallskip}
                      \sigma(m){\rm e}^{2\nu x}&z^{-1}}. \eqno(3.12)$$
This insures a proper adelic match with the $p$-adic cases treated in
Section 2.

With this choice, some straightforward calculations give the following
form for the $M$-matrix

$$\eqalignno{&\qquad M^m \sim &(3.13)\cr
&\pmatrix{\textfont0 = \eightrm \textfont1 = \eighti \scriptfont0 = \sixrm
\scriptfont1 = \sixi a(m,k)-{{b(m,k)}\over{2ik}}\sigma(m){\rm
e}^{2ikx}&\textfont0 = \eightrm \textfont1 = \eighti \scriptfont0 = \sixrm
\scriptfont1 = \sixi b(m,k){\rm
e}^{2(ik-\nu)x}\cr\noalign{\smallskip}\textfont0 = \eightrm \textfont1 =
\eighti \scriptfont0 = \sixrm \scriptfont1 = \sixi {\rm e}^{2\nu
x}\Bigl[{{\sigma(m+1)a(m,k)-\sigma(m)d(m,k)}
\over{2ik}}+c(m,k){\rm e}^{-2ikx}+{{b(m,k)}\over{4k^2}}{\rm
e}^{2ikx}\Bigr]&\textfont0 = \eightrm \textfont1 = \eighti \scriptfont0 =
\sixrm \scriptfont1 = \sixi
d(m,k)+{{b(m,k)}\over{2ik}}\sigma(m+1){\rm e}^{2ikx}}\cr} $$
where $a(m,k)$, $b(m,k)$, $c(m,k)$ and $d(m,k)$ all are $x$-independent. The
complete derivation of these results is given in Appendix B.
The ZS doublet at time $m$ then has the asymptotic form

$$\eqalign{u^m(x) &\sim u^0(x)
\prod_{j=0}^{m-1}{\Bigl[a(j,k)+{{b(j,k)\sigma(j)}\over{2ik}}{{c^j_-\,(k;\infty)}\over{c^j_+\,(k;\infty)}} \Bigr]}\cr
           v^m(x) &\sim  \tau(x)\Biggl[{\rm
e}^{(ik-1)x}\,c^0\,(k;\infty)\,{{\sigma(m)}\over{\sigma(0)}}\,\prod_{j=0}^{m-1}{\Bigl[a(j,k)+{{b(j,k)\sigma(j)}\over{2ik}}{{c^j_-\,(k;\infty)}\over{c^j_+\,(k;\infty)}} \Bigr]} + \cr
& \hskip 0.7in  + {\rm
e}^{-(ik+1)x}\,c^0\,(-k;\infty)\,\prod_{j=0}^{m-1}{\Bigl[d(j,k)+{{2ik\,c(j,k)}\over{\sigma(j)}}{{c^j_+\,(k;\infty)}\over{c^j_-\,(k;\infty)}} \Bigr]} \Biggr] \cr } \eqno(3.14)$$
where $c^j_+\,(k;\infty)$, $c^j_-\,(k;\infty) $ obey the recursion
relations

$$\eqalign{{{c^{j+1}_+\,(k;\infty)}\over{\sigma(j+1)}} &=
a(j,k)\,{{c^j_+\,(k;\infty)}\over{\sigma(j)}} +
b(j,k)\,{{c^j_-\,(k;\infty)}\over{2ik}}\cr
c^{j+1}_-\,(k;\infty) &= d(j,k)\,c^j_-\,(k;\infty) + 2ik\,
c(j,k)\,{{c^j_+\,(k;\infty)}\over{\sigma(j)}}\cr}\eqno(3.15)$$

This is precisely the real analog of the $p$-adic equation
(2.16). Again we enforce $S$-matrix unitarity, by imposing
$${{b(j,k)\sigma(j+1)}\over{2ik}} =
-{{2ik\,c(j,k)}\over{\sigma(j)}}\eqno(3.16a)$$
and
$$\sigma(j)\,d(j,k) = \sigma(j+1)\,a(j,k) \eqno(3.16b) $$
and, therefore, $[c^j_-\,(k;\infty)]^* = c^j_+\,(k;\infty)$.
Here again, $\sigma(m)$, $a(m,k)$, $b(m,k)$, $c(m,k)$ and $d(m,k)$ are
different from those
in section 2, as now they correspond to $p=\infty$. Then
$$ v^m(x) \sim \tau(x)\Bigl[{\rm e}^{(ik
-1)x}\,c^0(k;\infty)\prod_{j=0}^{m-1}{\Delta(j,k)} + {\rm
e}^{-(ik
+1)x}\,c^0(-k;\infty)\prod_{j=0}^{m-1}{\Delta^*(j,k)}\Bigr] \eqno
(3.14a)$$
where

$$\Delta(j,k) = d(j,k) \Bigl[1 +
2ik\,{{c(j,k)}\over{\sigma(j)d(j,k)}}{{c^j_+\,(k;\infty)}\over{c^j_-\,(k;\infty)}}\Bigr]\eqno(3.17)$$
and we see that
$$c^m(k;\infty) = c^0(k;\infty)\,\Lambda(m,k\,;\infty) \eqno(3.18)$$
where
$$\Lambda(m,k\,;\infty) = \prod_{j=0}^{m-1}{\Delta(j,k)} \eqno(3.19)$$

As we shall see in the next section this has precisely the right form
for an adelic product formula.

The asymptotics of $u^m(x)$ yields the $p=\infty$ counterpart of the
$l^m$ of section 2. The same $\Lambda(m,k;\infty)$ function (3.19) is involved.
Everything parallels the $p$-adic case, as expected.

For completeness, in Appendix C we discuss the integrable evolution of
the real zsf problem with ZS system of type (3.2).

\newsec {Integrable Evolution of the Adelic Zonal Spherical Function Problem}
\bigskip

In sections 2 and 3 we treated an infinite set of ZS problems, one for each
Bruhat-Tits tree with finite prime branching number $p$ and one for the ZS
problem on the
real hyperbolic plane, the case $p=\infty$. We found similar evolutions in
discrete time $m$ for, the Jost-Harish-Chandra $c$-functions in all these
cases. Specifically
$$
c^{m}(k;v)= c^{0}(k;v)\, \Lambda(m,\chi^{m}(v),k;v)  \eqno(4.1)
$$
 with $c^{0}(k;v)$ the ordinary Jost-Harish-Chandra $c$-function for the
local hyperbolic planes (following arithmetic usage, $v$ labels the places of
the field {\an Q} of rational numbers and runs over all finite primes and the
infinite prime $v=\infty$, which denotes the place at which completion
of the rationals yields the ordinary real numbers) and
$$
c^{0}(k;v)={{\zeta_{v}(ik)}\over{\zeta_{v}(ik+1)}}, \eqno(4.2a)
$$
$$
\zeta_{p}(ik)= {{1}\over{1-p^{-ik}}}, \eqno(4.2b)
$$
$$
\zeta_{\infty}(ik)= \pi^{-{{ik}\over{2}}} \Gamma({{ik}\over{2}}),
\eqno(4.2c)
$$
$$
\Lambda(m,\chi^{m},k;p)= \prod_{j=0}^{m-1}{[1- \chi^{m}(p) p^{-ik}]},
\eqno(4.2d)
$$
$$
\Lambda(m,\chi^{m},k;\infty)=\prod_{j=0}^{m-1}{\Delta (j,k)}.
 \eqno(4.2e)
$$

One passes from these local evolutions to the adelic evolution, by performing
the adelic product over all the finite and infinite places $v$. At time $m=0$
this yields the familiar result
$$
c^{0}_{\ai A}(k)= {{\zeta_{\ai A}(ik)}\over{\zeta_{\ai A}(ik+1)}}, \eqno(4.3a)
$$
 where
$$
\zeta_{\ai A}(s)= {\pi^{-{{s}\over{2}}}} \Gamma({{s}\over{2}}) \zeta(s),
\eqno(4.3b)
$$
  $\zeta(s)$ being the ordinary Riemann zeta function and $\zeta_{\ai A}(s)$
the
adelic zeta function which obeys the simple functional equation
$$
\zeta_{\ai A}(1-s)= \zeta_{\ai A}(s).  \eqno(4.4)
$$

At time $m=0$, an interesting {\it adelic} problem (see \b) thus unifies all
the {\it local} problems considered in sections 2 and 3. Does this adelic
unification persist in the course of the time evolution? In general it does
not, but one can constrain the $p$-dependence of the integration constants
which appear in equation (4.2), in such a manner that the adelic unification
remain meaningful at {\it all times} $m$. By inspecting equation (4.2), it
becomes
evident that the $p$-dependence of the $\chi^{m}(p)$ must be such as to
allow an Euler product to be formed at each time $m$. The simplest way to
insure this
is, to fix, at each time $m$, the $p$-dependence of $\chi^{m}(p)$ to be that
of a Dirichlet character modulo some integer $r_m$, which can depend on the
discrete time $m$. Then the adelic $\Lambda$-function,
$$
\Lambda_{\ai A}(m, \chi ,k)= \prod_{v}{\Lambda(m,\chi^{m}(v),k;v)},
\eqno(4.5)
$$
 the product of all local $\Lambda$-functions of equation
(4.2), evolves in a simple fashion:
$$
\Lambda_{\ai A}(m+1, \chi ,k)= L_{\ai A}(\chi^{m},ik)\, \Lambda_{\ai A}(m, \chi
,k) \eqno(4.6)
$$
with $L_{\ai A}$ the adelic Dirichlet $L$-function corresponding to the
Dirichlet character $\chi^{m}$:
$$
L_{\ai A}(\chi^{m},ik)=L_{\infty}(\chi^{m},ik)\, L(\chi^{m}, ik). \eqno(4.7)
$$
Here
$$
L(\chi^{m}, ik)=\prod_{p}{{1}\over{1-\chi^{m}(p)}} \eqno(4.8)
$$
 is the Dirichlet $L$-function corresponding to the character $\chi^m$ and
$L_{\infty}(\chi, ik)$ its gamma factor. Specifically {\n}, if the exponent
$\epsilon$ of the character $\chi^m$ is defined by
$$
(-1)^{\epsilon_m}={{\chi^{m}(-1)}\over{\chi^{m}(1)}},  \eqno(4.9)
$$
then
$$
L_{\infty}(\chi^{m}, ik)=
({{r_m}\over{\pi}})^{{ik}\over{2}}\, \Gamma({{ik+ \epsilon_m}\over{2}})
\eqno(4.10)
$$

It is now evident that the time evolution in the adelic case amounts to the
accretion at time $m$ of a factor $L_{\ai A}(\chi^{m},ik)$ by the adelic
Jost-Harish-Chandra $c$-function.

A similar adelic treatment can also be given to the other component, the
$u$-component of the ZS doublet, more precisely to the $l^m$ function
encoded in it. From the local $l^m$ functions (2.24) we can construct
an adelic $l^m$ function, by forming the Euler product and including the
place at infinity. This adelic $l$-function then keeps accreting the same
$L_{\ai A}(\chi^{m},ik)$ factors as the $c^m$-function. The evolution of
the adelic ZS system is thus fully (i.e. for both components of the ZS
doublet) determined by a single  adelic $\Lambda_{\ai A}$ function.

All this can be considerably generalized,
by weakening the condition that an adelic amplitude be obtained at {\it all
times} $m$. If instead, we only require an adelic amplitude at all {\it
even} values of the discrete time $m$, as if though time steps of the adelic
system were twice longer than those of the local systems, then Langlands-type
$L$-functions for $GL(2)$ can be accreted. To see how this comes about, let
us consider a cusp form $f$ of weight $k$, which is an eigenfunction of all
Hecke
operators $T(p)$. Let $a_n$ be the Fourier coefficients of $f$ and let $a_1=1$.
Then the local Langlands $L$-functions corresponding to $f$ are
$$
L_p(s,f)={{1}\over{1-a_{p} \,p^{-s}+p^{k-1-2s}}}=
{{1}\over{(1-\mu_p \,p^{-s+{{k-1}\over{2}}})(1-\nu_p
\,p^{-s+{{k-1}\over{2}}})}}
 \eqno(4.11a)
$$
 with
$$
\mu_p \,\nu_p=1 \qquad \mu_p+ \nu_p= a_p \,p^{{1-k}\over{2}},
\eqno(4.11b)
$$
 so that setting
$$
\chi^{2j+1}(p)= \mu_p \,p^{{k-1}\over{2}}  \eqno(4.12a)
$$
 and
$$
\chi^{2j+2}(p)= \nu_p \,p^{{k-1}\over{2}},  \eqno(4.12b)
$$
 we pick up a $GL(2)$ Langlands $L$-factor at each positive {\it even} value
of the discrete time $m$. Similarly requiring the adelic time to correspond to
every third, fourth or higher step of the ordinary ``local'' time, we can
accrete higher Langlands $L$-functions ($GL(3)$,...), Shimura symmetric
squares of $L$-functions, etc...

It can also happen that the $L$-function accretion is irregular, say at time
$m=1$ we accrete a Dirichlet $L$-function, then at time $m=3$ a $GL(2)$
Langlands $L$-function, then at times $m=4,5,6$ again Dirichlet $L$-functions,
and so on without any visible Dirichlet-Langlands-Shimura-... pattern. All the
local $\chi^{m}(p)$ at each time $m$ are given in terms of the {\it arbitrary}
local parameters of the time evolution ($d^{m}(p)$ and $\delta^{m}(p)$
in equation
(2.21.b)). So, there is no mechanism directing the arithmetic evolution of the
system. In particular the adelizability at all times was obtained by
restricting the integration constants and not by explicit dynamical
constraints.
What we have found, is an infinite family of integrable systems, which
together give rise to adelic integrable systems with the just explained degree
of arbitrariness.

\newsec{Conclusions}

We have studied integrable ZS systems which for one of the components of the
ZS doublet, reduce at the initial time to the zsf problem on a (local) real
or $p$-adic hyperbolic plane. We have found that it is possible to so
coordinate the integrable evolutions of these systems, that at all later
times as well, meaningful adelic Jost functions are obtained. These adelic
Jost functions involved $L$-functions of various kinds, Dirichlet,
Langlands, Shimura... Though the appearance of these number-theoretic
functions is interesting in its own right, it is far from fully understood.
First of all, it is not clear what replaces at later times the adelic
symmetric space on
which the adelic scattering problem at the initial time is defined. Moreover,
as we saw, there is a lot of freedom in the order in which the various types
of $L$-functions get accreted at later times. It would be interesting if a
dynamical principle could be found to determine the "arithmetic evolution"
of the system. This principle would ultimately have to account for the
arithmetically meaningful $p$-dependence of the $\chi^{m}(p)$ assumed in
section 4 to make an adelic evolution possible.

\eject
{\bf Appendix A. Derivation of the solution (2.15)}

In this appendix we find the solution for the time evolution (2.13) of the ZS
system (2.12), when the $z$-dependence of the $M$-matrix elements is
as given in eq.(2.11). Inserting (2.14) into (2.13) we obtain

$$\def\\#1#2 {^#1_#2}
 \eqalign{a^m_n + Q^{m+1}_n c^m_n & = a^m_{n+1} + P^m_n b^m_{n+1} \cr
          \alpha^m_n + Q^{m+1}_n\gamma ^m_n & = \alpha^m_{n+1} + P^m_n
\beta^m_{n+1} \cr
          \delta^m_n + P^{m+1}_n b^m_n & = \delta^m_{n+1} + Q^m_n c^m_{n+1} \cr
          d^m_n + P^{m+1}_n\beta ^m_n & = d^m_{n+1} + Q^m_n \gamma^m_{n+1} \cr
          \beta^m_n + Q^{m+1}_n d^m_n  & = b^m_{n+1} + Q^m_n a^m_{n+1}\cr
          c^m_n + P^{m+1}_n a^m_n  & = \gamma^m_{n+1} + P^m_n d^m_{n+1} \cr}
\qquad
\eqalign{   b^m_n + Q^{m+1}_n \delta^m_n & = 0 \cr
            \gamma^m_n + P^{m+1}_n \alpha^m_n & = 0 \cr
            \beta^m_n + Q^m_{n-1} \alpha^m_n & = 0 \cr
            c^m_n + P^m_{n-1} \delta^m_n & = 0 \cr }  \eqno({\rm A}.1)$$

{}From these equations, one readily finds

$$\def\\#1#2 {^#1_#2}
   \eqalign{a\\mn & = Q^{m+1}_nP^m_{n-1}\delta\\mn +  a(m) \cr
           b\\mn & = -Q^{m+1}_n\delta\\mn \cr
           c\\mn & = -P^m_{n-1}\delta\\mn \cr
           d\\mn & = Q^m_{n-1}P^{m+1}_n\alpha\\mn +  d(m) \cr }
\qquad
\eqalign{\beta\\mn & = -Q^m_{n-1}\alpha\\mn \cr
         \gamma\\mn & = -P^{m+1}_n\alpha\\mn \cr}  \eqno({\rm A}.2a) $$

$$ \alpha^m_n = \Biggl[
\prod_{i=0}^{n-1}{{\bigl(1-Q^{m+1}_iP^{m+1}_i\bigr)}\over{\bigl(1-Q^m_nP^m_n
\bigr)}}\Biggr] \alpha(m) \quad  \delta^m_n = \Biggl[
\prod_{i=0}^{n-1}{{\bigl(1-Q^{m+1}_iP^{m+1}_i\bigr)}\over{\bigl(1-Q^m_nP^m_n
\bigr)}} \Biggr] \delta(m) \eqno({\rm A}.2b)$$

\noindent and two coupled equations for $Q^m_n$ and $P^m_n$:

$$\eqalign{ Q^m_n  a(m) &-Q^{m+1}_n  d(m) =  \cr
& \Biggl[
\prod_{i=0}^{n-1}{{\bigl(1-Q^{m+1}_iP^{m+1}_i\bigr)}\over{\bigl(1-Q^m_iP^m_i
\bigr)}}\Biggr]\bigl(1-Q^{m+1}_nP^{m+1}_n\bigr)\bigl(Q^{m+1}_{n+1}\delta(m) -
Q^m_{n-1}\alpha(m)\bigr)\cr} \eqno({\rm A}.3a)$$
$$\eqalign{ P^m_n  d(m) &-P^{m+1}_n  a(m) =  \cr
&           \Biggl[
\prod_{i=0}^{n-1}{{\bigl(1-Q^{m+1}_iP^{m+1}_i\bigr)}\over{\bigl(1-Q^m_iP^m_i
\bigr)}}\Biggr]\bigl(1-Q^{m+1}_nP^{m+1}_n\bigr)\bigl(P^{m+1}_{n+1}\alpha(m) -
P^m_{n-1}\delta(m)\bigr) \cr} \eqno({\rm A}.3b)$$

\noindent where $a(m)=a^m_0$, $\alpha(m)=\alpha^m_0$, $d(m)=d^m_0$ and
$\delta(m)=\delta^m_0$ are arbitrary
functions depending on the discrete time $m$ and on the lattice
branching number $p$,
but we choose not to show the latter dependence explicitly.

Dividing the two equations (A.3), we have

$$\eqalign{\alpha(m) a(m)\bigl(Q^m_{n-1}P^{m+1}_n - Q^m_n P^{m+1}_{n+1}\bigr) +
  \delta(m) d(m)\bigl(Q^{m+1}_{n+1}P^m_n - Q^{m+1}_nP^m_{n-1}\bigr) + & \cr
  +\alpha(m) d(m)\bigl(Q^{m+1}_nP^{m+1}_{n+1} - Q^m_{n-1} P^m_n\bigr) +
  \delta(m) a(m)\bigl(Q^m_nP^m_{n-1} - Q^{m+1}_{n+1}P^{m+1}_n\bigr)& = 0\cr}
\eqno({\rm A}.4) $$

{}From eq.(A.4)  we recognize the  particular solution:

$$Q^{m+1}_{n+1} ={ \sigma_p(m)\over \sigma_p(m+1)} Q^m_n \qquad  P^{m+1}_{n+1}
=P^m_n{\sigma_p(m+1)\over\sigma_p(m)}\eqno({\rm A}.5a)$$
with
$$\bigl[\delta(m) d(m) ({\sigma_p(m)})^2 -\alpha(m) a(m)
({\sigma_p(m+1)})^2\bigr]  \bigl(Q^m_{n-1}\,P^m_{n-1}-Q^m_n\,P^m_n\bigr)=0
\eqno({\rm A}.5b)  $$

There might be some further solutions of this type, but those would require
that $Q^m_n$ and $P^m_n$ also satisfy some supplementary conditions (such as
$P^m_{n-1}=  P^m_{n+1}$, for example), which have to be reflected at time
$m=0$, and which are not satisfied in our case, so we will not concern
ourselves with them.

Since  $n \ge 0$ and $Q^0_n$ vanishes in our case, (see (2.8)), this  solution
will produce a $Q^m_n $ which vanishes for $n > m$ at any finite time $m$. In
other words $Q^m_n = 0 $ asymptotically at all times.
As we are interested in the Jost functions, i.e. the asymptotic
scattering regime, this means that for our purposes $Q^m_n$ can be set
to zero at all times $m$ and all radial distances $n$, $Q^m_n = 0 $.
 A nice feature of such a choice
 of solution is that it preserves exactly, not only asymptotically, the
$p$-adic zsf-problem structure for the equation describing $v^m_n$ at
all times.

Imposing then
$$Q^m_n = 0 \eqno({\rm A.6a})$$
eqs. (A.5) are solved by
$$P^m_n = \sigma_p(m)p^{\rho(n-m)} \eqno({\rm A.6b})$$

This yields precisely the $R$-matrix of eq. (2.15a). Inserting eqs.(A.6)
into eqs.(A.2) yields the $M$-matrix (2.15b). Once we impose
(A.6a), eq.(A.5b) becomes an identity. So far we have only used
eq.(A.4) which is the ratio of the two eqs.(A.3). Eq.(A.3a) is now an
identity, too, and we thus require that eq.(A.3b) be obeyed. Inserting
eqs.(A.6) into (A.3b) one immediately finds eq.(2.15c). We thus showed
that the equations (2.15) do indeed solve the consistency equation (2.13).

\eject
{\bf Appendix B. Derivation of equation (3.13)}

We will find here an exact solution for the time-evolution equation
(3.11) of the real ZS system (3.2'). The form of the $M-$matrix is
the same as in the previous paragraph

$$M^m(x) = \pmatrix {A^m(x,k)& B^m(x,k)\cr \noalign{\smallskip} C^m(x,k)&
D^m(x,k)\cr}. \eqno({\rm B}.1)$$

With the choice (3.12) for the $R-$matrix, equation (3.11) produces the
following equations
$$\eqalign{ (A^m)' & = - B^m\sigma(m)\,{\rm e}^{2\nu x}\cr
(B^m)' & = B^m\,(z-z^{-1})\cr
(C^m)' & = -C^m\,(z-z^{-1}) +{\rm e}^{2\nu x}\bigl[A^m\sigma(m+1) -
D^m\sigma(m)\bigr]  \cr
(D^m)' & =  B^m\sigma(m+1)\,{\rm e}^{2\nu x}\cr} \eqno({\rm B}.2)$$

It is readily seen that the second equation of this system yields

$$B^m(x,k) = b(m,k)\, {\rm e}^{2(ik-\nu)x} \eqno({\rm B}.3)$$
which, when used in the first and fourth equation gives in turn

$$A^m(x,k) = a(m,k) - {{b(m,k)}\over{2ik}}\sigma(m){\rm e}^{2ikx}
\eqno({\rm B}.4)$$
and

$$D^m(x,k) = d(m,k) + {{b(m,k)}\over{2ik}}\sigma(m+1){\rm e}^{2ikx}.
\eqno({\rm B}.5)$$
Finally a straightforward integration of the third equation in
the system (B.2) results in

$$\eqalign{C^m(x,k) &=  {{b(m,k)}\over{4k^2}}\sigma(m)\sigma(m+1)\,{\rm
e}^{2(ik+\nu)x}  + c(m,k)\,{\rm e}^{2(\nu-ik)x} + \cr &
+{{\sigma(m+1)a(m,k)-\sigma(m)d(m,k)}
\over{2ik}}\,{\rm e}^{2\nu x}\cr} \eqno({\rm B}.6)$$
where all the 'constants of integration' $a(m,k), b(m,k), c(m,k)$, and
$d(m,k)$ are arbitrary functions of $z$, and implicitly of $k$,(see
equation (3.8)).

\eject
{\bf Appendix C. The ZS system (3.2)}

Let us now return to the Z.S. system (3.2) mentioned at  the beginning of
section 3.

Just like in the 'exact' case treated in section 3, we get a second
order differential equation for $v^0$
$$(v^0)'' -{{(P^0)'}\over{P^0}}(v^0)' + \Biggl[\zeta^2 -i\zeta
{{(P^0)'}\over{P^0}} - Q^0P^0\Biggr]v^0 = 0 \eqno({\rm C}.1)$$

Again $\zeta$ is related to the spectral parameter $k$ of equation (3.1). The
gauge
transformation

$$v^0(x) = \rho (x)t^0(x). \eqno({\rm C}.2)$$
and the same logic that led to equations (3.6), now yields
$$Q^0(x) = {\sigma(0)\over{\sinh^2 2x}}\, {\rm e}^{-2\nu x} \quad {\rm and}
\quad
  P^0(x) = -{1\over{\sigma(0)}}\,{\rm e}^{2\nu x} \eqno({\rm C}.3) $$
with
$$k = \pm \bigl(\zeta -i\nu \bigr) \quad {\rm and}\quad \rho (x) =
\rho_0 {\rm e}^{\nu x}\sqrt{\sinh 2x} \eqno({\rm C}.4) $$
Using the known solution (3.9), (3.10) to (3.1),
we  now look for an integrable time evolution of the system which at
all times  is
compatible with the zsf equation (3.1). That is, we wish to find a
solution to
equation (3.11), but, this time, by expanding in $\zeta$.
Again, because we want to preserve as much as possible the form of the
$R^m$-matrix, we will make
the following assumptions:

$$Q^m(x) = {\sigma(m)\over{\sinh^2 2x}}\, {\rm e}^{-2\nu x} \quad {\rm and}
\quad
  P^m(x) = \tau(m)\,{\rm e}^{2\nu x} \eqno({\rm C}.5) $$

These assumptions together with the expansion

$$ \eqalign{A^m(x,\zeta) = a(m,x)\zeta +
\alpha(m,x)\qquad& B^m(x,\zeta) = b(m,x)\zeta + \beta(m,x)\cr
C^m(x,\zeta) = c(m,x)\zeta + \gamma(m,x) \qquad & D^m(x,\zeta) = d(m,x)\zeta +
\delta(m,x) \cr } \eqno({\rm C}.6)$$
lead to the system:
$$\eqalign{a^m(x)' &= \sigma(m+1)c^m(x)\,{{{\rm e}^{-2\nu x}}\over{\sinh^2 2x}}
- \tau(m)b^m(x)\,{\rm e}^{2\nu x}\cr
d^m(x)' &=  \tau(m+1)b^m(x)\,{\rm e}^{2\nu x} - \sigma(m)c^m(x)\,{{{\rm
e}^{-2\nu x}}\over{\sinh^2 2x}}\cr
\alpha^m(x)' &= \sigma(m+1)\gamma^m(x)\,{{{\rm e}^{-2\nu x}}\over{\sinh^2 2x}}
- \tau(m)\beta^m(x)\,{\rm e}^{2\nu x}\cr
\delta^m(x)' &=  \tau(m+1)\beta^m(x)\,{\rm e}^{2\nu x} -
\sigma(m)\gamma^m(x)\,{{{\rm e}^{-2\nu x}}\over{\sinh^2 2x}}\cr
2i\beta^m(x) &= \bigl[a^m(x)\sigma(m) - d^m(x)\sigma(m+1)\bigr]\,{{{\rm
e}^{-2\nu x}}\over{\sinh^2 2x}}\cr
2i\gamma^m(x) &= \bigl[a^m(x)\tau(m+1) - d^m(x)\tau(m)\bigr]\,{\rm e}^{2\nu
x}\cr
\beta^m(x)' &= \bigl[\delta^m(x)\sigma(m+1)-\alpha^m(x)\sigma(m)\bigr]\,{{{\rm
e}^{-2\nu x}}\over{\sinh^2 2x}}\cr
\gamma^m(x)' &= \bigl[\alpha^m(x)\tau(m+1) - \delta^m(x)\tau(m)\bigr]\,{\rm
e}^{2\nu x}\cr b^m(x) &= 0 \quad  c^m(x) = 0\cr}\eqno({\rm C}.7) $$

In a straightforward manner, the solution for this system is found to
be
$$\eqalign{a^m(x) = 2ia(m)\quad & \alpha^m(x) =
-{{a(m)}\over2}\bigl[\tau(m+1)\sigma(m+1) -\tau(m)\sigma(m)\bigr]\,\coth 2x +
\alpha(m)\cr
d^m(x) = 2id(m)\quad & \delta^m(x) = {{d(m)}\over2}\bigl[\tau(m+1)\sigma(m+1) -
\tau(m)\sigma(m)\bigr]\,\coth 2x + \delta(m)\cr
b^m(x) = 0 \quad & \beta^m(x) = \bigl[a(m)\sigma(m) -
d(m)\sigma(m+1)\bigr]\,{{{\rm e}^{-2\nu x}}\over{\sinh^2 2x}}\cr
c^m(x) = 0 \quad & \gamma^m(x) = \bigl[a(m)\tau(m+1) - d(m)\tau(m)\bigr]\,{\rm
e}^{2\nu x}\cr}\eqno({\rm C}.8)$$
where the 'constants of integration' $a(m), d(m), \alpha(m)$ and
$\delta(m)$ are subject to either the compatibility conditions

$$\eqalign{\alpha(m)\tau(m+1) - \delta(m)\tau(m) &= 2\nu\bigl[a(m)\tau(m+1) -
d(m)\tau(m)\bigr] \cr
a(m)\tau(m+1) + d(m)\tau(m) &= 0 \cr
\sigma(m+1)\delta(m) -\sigma(m)\alpha(m) &= -2\nu \bigl[a(m)\sigma(m) -
d(m)\sigma(m+1)\bigr]\cr
\bigl[\tau(m+1)\sigma(m+1) - \tau(m)\sigma(m)\bigr]^2 &=
8\bigl[\tau(m+1)\sigma(m+1) + \tau(m)\sigma(m)\bigr]\cr }\eqno({\rm C}.9)$$
{\it or} to the conditions
$$\eqalign{\tau(m+1)\sigma(m+1) &= \tau(m)\sigma(m)\cr
{{a(m)}\over{d(m)}} = {{\alpha(m)}\over{\delta(m)}} &=
{{\sigma(m+1)}\over{\sigma(m)}}\cr.}\eqno({\rm C}.10)$$

{}From the first three equations (C.9) we have
$$\eqalign{d(m) &=  - a(m){{\tau(m+1)}\over{\tau(m)}}\cr
\alpha(m) &=  2\nu a(m)\cr
\delta(m) &= -2\nu a(m){{\tau(m+1)}\over{\tau(m)}}\cr}\eqno({\rm C}.11)$$
whereas the last one determines the time-evolution of the $R$-matrix:
$$\bigl[\pi(m+1) - \pi(m)\bigr]^2 = 8 \Bigl[\pi(m) + \pi(m+1)\Bigr] \eqno({\rm
C}.12) $$
where we used the notation $\pi(m) =\sigma(m)\,\tau(m)$. Notice that
$d(m)$, $\alpha(m)$, and $\delta(m)$ are all proportional to
$a(m)$. Because of equation (C.8) so are then $\beta(m)$ and
$\gamma(m)$. Thus $a(m)$ is  an overall time-dependent normalization
of the time evolution matrix $M^m$. Without any loss of generality we
henceforth set $a(m)$ constant, say $a(m) = 1/2$.

We therefore start with equation (C.12) and introduce the function
$l(m)$ by

$$\pi(m) = l^2(m) - 1 \eqno({\rm C}.13) $$
Then equation (C.12) becomes a quadratic equation for $l(m+1)^2$ for a
given $l(m)$. Its two solutions are
$$l^2(m+1) = \bigl[l(m) \pm 2 \bigr]^2 \eqno({\rm C}.14)$$
The $R$-matrix at time $m=0$ determines $\pi(0)=-1$ so that
$$l(0) = 0. \eqno({\rm C}.15) $$

The corresponding solutions of (C.9), written in matrix form are:

$$R^{m+1}_\pm = \pmatrix{ik -\nu&{\sigma(m+1)\over{\sinh^2 2x}}\, e^{-2\nu
x}\cr \noalign{\smallskip} {[l(m) \pm 1][l(m)\pm 3]\over{\sigma(m+1)}}\,e^{2\nu
x}&-ik +\nu\cr}\eqno({\rm C}.16)$$
and
$$M^m_\pm = 2\,a(m) \pmatrix{\textfont0 = \eightrm \textfont1 = \eighti
\scriptfont0 = \sixrm \scriptfont1 = \sixi ik-\bigl[1\pm l(m)\bigr]\coth
2x&{\sigma(m)\over{\sinh^2r 2x}}\, e^{-2\nu x}{{l(m)\pm 1}\over{l(m) \mp
1}}\cr\noalign{\smallskip}{{[l(m) \pm 1][l(m) \pm
3]\over{\sigma(m+1)}}\,e^{2\nu x}}
&\textfont0 = \eightrm \textfont1 = \eighti \scriptfont0 = \sixrm \scriptfont1
= \sixi {-\bigl[ik+\bigl[1\pm l(m)\bigr]\coth
2x\bigr]{{\sigma(m)}\over{\sigma(m+1)}}{{l(m) \pm 3}\over{l(m) \mp 1}}}\cr}.
\eqno({\rm C}.17)$$

For the alternate set (C.10) of compatibility conditions one finds

$$\pi(m+1) = \pi(m)  \eqno({\rm C}.18)$$
and the corresponding solutions
$$R^{m+1} =R^m \quad{\rm and}\quad M^m = \pmatrix{a(m)\bigl(ik -\nu\bigr)&0\cr
0&d(m)\bigl(ik -\nu\bigr)\cr}. \eqno({\rm C}.19)$$

 Thus, at any time there are  five  possible evolutions, the four (due to
the {\it two} sign ambiguities in the quadratic equation (C.14),)

$$l(m) \rightarrow \, l(m+1)_\pm \,= \pm l(m) \pm 2  \eqno({\rm C}.20)$$
and the solution (C.18), (C.19).
For simplicity we  choose the time evolution corresponding to $l(m)
\rightarrow l(m) + 2$ so as to avoid 'returns', 'reflections',
or 'stagnations'. Then $l(m) = 2m$ and the asymptotic solution of the Z-S
system is

$$ \eqalignno{u^m(x) =& -2ik \sigma(0)\,e^{-\nu x}\sqrt{\sinh
2x}\,c^m(k)\,e^{(ik - 1)x} &({\rm C}.21)\cr
v^m(x) =& \,\bigl(2m-1\bigr)\bigl(2m+1\bigr){{\sigma(0)}\over{\sigma(m)}}e^{\nu
x}\sqrt{\sinh 2x}\,\Bigl[c^m(k)\,e^{(ik - 1)x} +c^m(-k)\,e^{(-ik -
1)x}\Bigr]\cr} $$

The Jost functions at time $m$ are given by expressions that are
similar to those  we obtained for the $p$-adic case:

$$c^m(k) = c^0(k;\infty)  \prod_{l=1}^m
\bigl[ik-\bigl(2l-1\bigr)\bigr] \eqno({\rm C}.22)$$
with $c^0(k;\infty)$ given by equation (3.10), and the conjugate
relation for $c^m(-k)$. Although the form of (C.22) is similar to the
one in (2.22),  it is not suited for adelization.  This is evident
from the appearance of the factors linear in  $K$ as opposed to the
expected gamma factors. Presumably this is due to the expansion in
$\zeta$, undertaken here, which is different from the $z$-expansion in
section 3. It is amusing to note that the linear factors in (C.22)
are themselves ``gamma-like'' functions of one order lower in the
Barnes hierarchy {\5}.

We note that the second order equation obeyed by  $w^m(x)$
can be brought to the form

$$(w^m)''+ 2\coth 2x \,(w^m)' + \Biggl(k^2+1-{4m^2\over{\sinh^2 2x}}\Biggr)
\,w^m = 0 \eqno({\rm C}.23) $$
\noindent or, equivalently, with $w^m(x) = {z(x)\over{\sqrt{\sinh 2x}}}$

$$z''+\Biggl(k^2-{{4m^2-1}\over{\sinh^2 2x}}\Biggr) \,z = 0 \eqno({\rm C}.24)
$$
which is the eigenvalue problem for the $P^m_{k}(x)$, i.e. the associated
Legendre functions \a.

\vskip 1.5cm
{\bf Acknowledgement}

\bigskip
The author is very much indebted to Prof. Peter Freund for
suggesting the theme, and for  many fruitful and edifying
discussions.

\listrefs

\end{document}